\documentclass[
	aps,
	amsmath,amssymb,
	reprint,
	superscriptaddress,
	pra
   ]{revtex4-2}

\usepackage{graphicx}% Include figure files
\usepackage{color}
\usepackage{dcolumn}% Align table columns on decimal point
\usepackage{siunitx}
\usepackage{bm}% bold math
\usepackage{braket}
\usepackage{physics}
\usepackage{url}
\usepackage[colorlinks=true,citecolor=blue,linkcolor=magenta]{hyperref}
\usepackage{cleveref}
\crefname{equation}{Eq.}{Eqs.}
\crefname{figure}{Fig.}{Figs.}
\crefname{section}{Sec.}{Secs.}
\begin{document}

%\preprint{AIP/123-QED}

\title[]{Quantum-inspired algorithm applied to extreme learning}

\author{Iori Takeda}
\affiliation{Graduate School of Engineering Science, Osaka University, 1-3 Machikaneyama, Toyonaka, Osaka 560-8531, Japan.}

\author{Souichi Takahira}
\email{takahira.souichi.es@osaka-u.ac.jp}
\affiliation{Graduate School of Engineering Science, Osaka University, 1-3 Machikaneyama, Toyonaka, Osaka 560-8531, Japan.}

\author{Kosuke Mitarai}
\email{mitarai.kosuke.es@osaka-u.ac.jp}
\affiliation{Graduate School of Engineering Science, Osaka University, 1-3 Machikaneyama, Toyonaka, Osaka 560-8531, Japan.}
\affiliation{Center for Quantum Information and Quantum Biology, Institute for Open and Transdisciplinary Research Initiatives, Osaka University, Japan.}

\author{Keisuke Fujii}
\email{fujii.keisuke.es@osaka-u.ac.jp}
\affiliation{Graduate School of Engineering Science, Osaka University, 1-3 Machikaneyama, Toyonaka, Osaka 560-8531, Japan.}
\affiliation{Center for Quantum Information and Quantum Biology, Institute for Open and Transdisciplinary Research Initiatives, Osaka University, Japan.}
\affiliation{Center for Quantum Computing, RIKEN, Wako Saitama 351-0198, Japan.} 

\date{\today}

\begin{abstract}
	
Quantum-inspired singular value decomposition (SVD) is a technique to perform SVD in logarithmic time with respect to the dimension of a matrix, given access to the matrix embedded in a segment-tree data structure.
The speedup is possible through the efficient sampling of matrix elements according to their norms.
Here, we apply it to extreme learning which is a machine learning framework that performs linear regression using random feature vectors generated through a random neural network.
The extreme learning is suited for the application of quantum-inspired SVD in that it first requires transforming each data to a random feature during which we can construct the data structure with a logarithmic overhead with respect to the number of data.
We implement the algorithm and observe that it works order-of-magnitude faster than the exact SVD when we use high-dimensional feature vectors.
However, we also observe that, for random features generated by random neural networks, we can replace the norm-based sampling in the quantum-inspired algorithm with uniform sampling to obtain the same level of test accuracy due to the uniformity of the matrix in this case.
The norm-based sampling becomes effective for more non-uniform matrices obtained by optimizing the feature mapping.
It implies the non-uniformity of matrix elements is a key property of the quantum-inspired SVD.
This work is a first step toward the practical application of the quantum-inspired algorithm.
\end{abstract}

\pacs{Valid PACS appear here}
\maketitle

\section{Introduction}

In 2016, Kerenedis and Prakash have designed a quantum algorithm for preparing a low-rank approximation of an $n\times m$ matrix in $\mathrm{polylog}(nm)$ time on a quantum computer \cite{Kerenidis2016}.
They applied the technique to construct recommendation systems, which is thought to be one of quantum speedups for practical problems.
The key ingredient that makes the algorithm efficient was the segment-tree data structure.
Inspired by this work, Tang \cite{tang2018quantum} designed a ``quantum-inspired'' classical algorithm that also works in $\mathrm{polylog}(nm)$ by exploiting the data structure.
It utilizes the fact that the segment tree allows us to efficiently sample matrix elements according to probabilities proportional to their Frobenius norm. 
Motivated by Tang's breakthrough, quantum-inspired classical algorithms for other tasks, 
such as matrix inversion and linear regression, have been proposed 
\cite{NMF2019,LinEqu2020,Gilyen2022improvedquantum,majima2021qiCCA,SDP2020,qiSVT,FWdeQML2020}. % 追加(高比良)
Those algorithms also have polylogarithmic complexity in the matrix dimension if the low-rank approximation of the matrix is valid and the data structure is constructed in advance.  % 追加(高比良)

Although this quantum-inspired algorithm has polylogarithmic complexity in the matrix dimension, it is still not well-known that it works in reasonable runtime for practical tasks.
A possible bottleneck in practice is that it assumes the availability of the segment tree, which takes $O(nm\log(nm))$ time to construct in general.
We therefore must find an application where this cost for constructing data structure does not matter.
Also, the runtime analyses of previous works are rather pessimistic in that they have a very large constant prefactor.
However, there is a possibility that the prefactors are an artifact required to prove a rigorous theorem, and the overhead becomes smaller when using it in practice \cite{Arrazola2020quantuminspired}. % 参考文献を最後に追加(高比良)

Machine learning is a field where the low-rank approximation of matrices plays an important role. 
Several quantum-inspired algorithms for machine learning have been proposed \cite{Tang2021qiPCA,FWdeQML2020}. % 追加(高比良)
In this work, we apply the algorithm to machine learning via the framework called extreme learning \cite{huang2004extreme}.
In the extreme learning, we construct a model $f(\bm{x})$ by linear combination of randomly chosen features $\{\phi_i(\bm{x})\}_{i=1}^M$.
The training of the model is accomplished by computing a pseudo-inverse of a $D\times M$ matrix, where $D$ is the number of training data.
We seek to speed up this training process by using the quantum-inspired low-rank approximation algorithm.
The extreme learning is suited for the application of the quantum-inspired algorithm in the sense that we must preprocess the input data to the random features $\phi_i(\bm{x})$ for every training data, and thus $O(DM)$ computational cost is not avoidable in the first place.
The segment tree data structure can be constructed with an additional $O(\log(DM))$ cost, making the total preprocessing cost $O(DM\log(DM))$, which is a slight increase from the original cost.

We perform numerical experiments on two famous image datasets, MNIST handwritten digits \cite{MNIST} and CIFAR-10 \cite{CIFAR10}, using the quantum-inspired algorithm.
Our numerical experiments show that it can significantly reduce the time required for training without much program code optimizations compared to the approach based on the exact low-rank approximation.
On the other hand, we find that the weighted sampling of the matrix elements, which is the core idea of the quantum-inspired algorithm, is not required but we can use uniform sampling to achieve the same level of performance in the naive setting of extreme learning.
It is because the matrix elements become rather uniform due to the randomness of the feature $\phi_i(\bm{x})$.
To see more distinct advantages of using the quantum-inspired algorithm, we also conduct experiments using extreme learning with optimized features instead of the random features.
This approach makes the matrix elements non-uniform, and we can observe the advantage of using it.
The experiments show us that the quantum-inspired algorithm is effective in practical settings although certain care is needed to get the most out of it.
This work is a first step toward the practical application of the quantum-inspired algorithm.

\section{Theory}

\subsection{Quantum-inspired low-rank approximation}\label{sec:quantum-inspired}
We first give a brief description of the quantum-inspired low-rank approximation algorithm \cite{tang2018quantum}.
The task is to find a low-rank approximation of a matrix $\bm{X}\in\mathbb{R}^{m\times n}$ where $m\leq n$.
By singular value decomposition, $\bm{X}$ can be expressed as,
\begin{align}
    \bm{X} = \sum_{i=1}^m \sigma_i \bm{u}_i\bm{v}^{\mathrm{T}}_i,
\end{align}
where $\bm{u}_i$ and $\bm{v}_i$ is left- and right-singular vectors for a singular value $\sigma_i$.
We assume that $\sigma_1 \geq \sigma_2 \geq \cdots \geq \sigma_m$.
The quantum-inspired algorithm of Ref. \cite{tang2018quantum} seeks an approximation of $\bm{X}$ in the form of,
\begin{align}
    \tilde{\bm{X}} = \sum_{i=1}^K \tilde{\sigma}_i \tilde{\bm{u}}_i\tilde{\bm{v}}^{\mathrm{T}}_i,
\end{align}
where $\tilde{\sigma}_i\approx\sigma_i$, $\tilde{\bm{u}}_i\approx\bm{u}_i$, and $\tilde{\bm{v}}_i\approx\bm{v}_i$.
In other words, the output of the algorithm with an input matrix $\bm{X}$ is $\{\tilde{\sigma}_i, \tilde{\bm{u}}_i, \tilde{\bm{v}}_i\}_{i=1}^K$.

Let us describe the concrete algorithm.
We refer to this algorithm as the mod-FKV algorithm following Ref.~\cite{tang2018quantum} since it is a slightly modified version of Frieze, Kannan, and Vempala's algorithm \cite{frieze2004fast}.
Here, we denote $(i,j)$-element of $\bm{X}$ by $\bm{X}(i,j)$, $i$-th row vector of $\bm{X}$ by $\bm{X}(i,:)$, and $j$-th column vector of $\bm{X}$ by $\bm{X}(:,j)$.
\begin{enumerate}
    \item Sample row indices of $\bm{X}$ with probability
    \begin{align}
        \label{eq:fi}
        f_i=\frac{\|\bm{X}(i,:)\|^2}{\|\bm{X}\|_F^2},
    \end{align}
    where $\|\bm{X}\|_F$ represents the Frobenius norm of $\bm{X}$.
    One sample form this probability distribution can be drawn in time $O(\log(m))$ with the assumption that $\bm{X}$ is stored in a segment-tree data structure \cite{tang2018quantum}.
    Let the sampled indices $\{i_1,i_2,\dots,i_P\}$.
    \item For each $i_p$, sample column indices $\{j_1, j_2, \dots, j_P\}$ with probability
    \begin{align}
        \label{eq:gj}
        g_j=\frac{1}{P}\sum_{p=1}^P\frac{\|\bm{X}(i_p,j)\|^2}{\|\bm{X}(i_p,:)\|^2},
    \end{align}
    which takes $O(\log(n))$ per sample assuming the data structure.
    \item Define a matrix $\bm{W}\in\mathbb{R}^{P\times P}$ by 
    \begin{align}
        \bm{W}(p,q) = \frac{\bm{X}(i_p, j_q)}{P\sqrt{f_{i_p}g_{j_q}}},
    \end{align}
    and perform its singular value deomposition. We obtain $\bm{u}_i^{\prime}\in \mathbb{R}^P$, $\bm{v}_i^{\prime}\in \mathbb{R}^P$ and $\tilde{\sigma}_i\in \mathbb{R}$ such that 
    \begin{align}
        \bm{W} = \sum_{i=1}^P \tilde{\sigma}_i \bm{u}_i^{\prime}{\bm{v}_i^{\prime}}^{\mathrm{T}}.
    \end{align}
    We assume $\sigma_1\geq\sigma_2\geq\cdots\geq\sigma_P$.
    This can be done in $O(P^3)$ time.
    \item Define a matrix $\bm{S}\in\mathbb{R}^{P\times n}$ by 
    \begin{align}
        \bm{S}(p,:)=\frac{\bm{X}(i_p,:)}{\sqrt{Pf_p}},
    \end{align}
    and calculate $\tilde{v}_i$ and $\tilde{u}_i$ as follows:
    \begin{align}
        \tilde{\bm{v}}_i &= \frac{1}{\tilde{\sigma}_i}\bm{S}^{\mathrm{T}}\bm{v}_i^\prime\\
        \tilde{\bm{u}}_i &= \frac{1}{\tilde{\sigma}_i}\bm{X}\tilde{\bm{v}}_i.
    \end{align}
    The calculation of each $\tilde{\bm{v}}_i$ takes $O(nP)$ time and that of each $\tilde{\bm{u}}_i$ takes $O(mn)$ time.
    \item Return $\{\tilde{\sigma}_i, \tilde{\bm{u}}_i, \tilde{\bm{v}}_i\}_{i=1}^K$.
\end{enumerate}
In the above algorithm, the number of samples $P$ can be taken as $\mathrm{poly}(K, 1/\epsilon)$ to output $\{\tilde{\sigma}_i, \tilde{\bm{u}}_i, \tilde{\bm{v}}_i\}_{i=1}^K$ such that $\|\tilde{\bm{X}}-\bm{X}_K\|_F \leq \epsilon$ where $\bm{X}_K = \sum_{i=1}^K \sigma_i \bm{u}_i\bm{v}_i^{\mathrm{T}}$ is the exact rank-$K$ approximation of $\bm{X}$.
In the rest of this paper, the mod-FKV algortihm with parameters $K$ and $P$ is referred to as mod-FKV($K,P$).

Note that the above algorithm does not achieve the complexity $\mathrm{polylog}(nm)$ because of the cost required to construct $\tilde{\bm{u}}_i$ and $\tilde{\bm{v}}_i$.
The original paper \cite{tang2018quantum} avoids the explicit construction of these vectors, which is not required for certain applications, thus achieving the polylogarithmic scaling.
However, we allow $O(mn)$ cost because our target application, extreme learning, already requires $O(mn)$ preprocessing cost to transform data $\bm{x}$ to features $\phi_i(\bm{x})$ as we see in the next subsection.

\subsection{Extreme learning}

Extreme learning \cite{huang2004extreme} constructs a model by linear combination of random features $\{\phi_i(\bm{x})\}_{i=1}^M$.
The random features $\{\phi_i(\bm{x})\}_{i=1}^M$ is taken as outputs of a feed-forward neural network with random connections.
For example, we can take $\phi_i(\bm{x}) = g(\bm{a}_i^\mathrm{T}\bm{x}+b_i)$ where $g$ is an activation function such as ReLu, $\bm{a}_i$ and $b_i$ are random vector and bias, respectively.
In this sense, we refer to $M$ as the number of nodes.

Given a training data set $\{(\bm{x}_i, y_i)\}_{i=1}^D$ where $\bm{x}_i$ and $y$ are input and teacher data respectively, we try to find a weight vector $\bm{w}\in\mathbb{R}^{M}$ such that $\sum_{i=1}^D\left(\bm{w}^\mathrm{T}\bm{\phi}(\bm{x}_i)-y_i\right)^2$ is minimized.
Such $\bm{w}$ can be calculated by using Moore-Penrose pseudo-inverse $\bm{X}^+$ of a matrix $\bm{X}=(\bm{\phi}(\bm{x}_1)~\bm{\phi}(\bm{x}_2)~\cdots~\bm{\phi}(\bm{x}_D))$ as
\begin{align}\label{eq:optimal-weight}
    \bm{w} = \bm{X}^+ \bm{y},
\end{align}
where $\bm{y}=(y_1, ~ y_2, ~ \dots, ~ y_D)^\mathrm{T}$.

\subsection{Our proposal}
In this work, we exploit the mod-FKV algorithm for computing the optimal weight $\bm{w}$ by Eq.~\eqref{eq:optimal-weight}.
More concretely, we first compute the matrix $\bm{X}$ and store it in the segment-tree data structure.
Then, we use the algorithm in Sec. \ref{sec:quantum-inspired} to compute a $K$-rank approximation of $\bm{X}$.
Using the obtained $\{\tilde{\sigma}_i, \tilde{\bm{u}}_i, \tilde{\bm{v}}_i\}_{i=1}^K$, we approximate pseudo-inverse of $\bm{X}$ by,
\begin{align}\label{eq:approx-pseudo-inverse}
    \tilde{\bm{X}}^+ = \sum_{i=1}^K \frac{1}{\tilde{\sigma}_i} \tilde{\bm{u}}_i\tilde{\bm{v}}^{\mathrm{T}}_i,
\end{align}
and calculate the weight vector as $\bm{w}=\tilde{\bm{X}}^+\bm{y}$.

In the following subsections, we compare the proposed approach with a conventional approach that uses exact singular value decomposition.

\section{Experiments}

We test our idea with two famous image datasets: MNIST handwritten digits \cite{MNIST} and CIFAR-10 \cite{CIFAR10}.
Both of the datasets consist of images that are labeled into ten distinct classes.
An input data $\bm{x}_i$ consists of pixel values of images.
We first normalize the input values to have a minimum value of 0 and a maximum value of 1.
We encode a teacher datum by one-hot encoding, i.e., for each $\bm{x}_i$, we have ten teacher data $\{y_i^{(l)}\}_{l=1}^{10}$ such that $y_i^{(l)}=1$ if $\bm{x}_i$ belongs to label $l$ and $y_i^{(l)}=0$ otherwise.

The random features $\phi_i(\bm{x})$ is taken as $\phi_i(\bm{x})=g(\bm{a}_i^\mathrm{T}\bm{x}+b_i)$ where $g$ is ReLu function, each element of $\bm{a}_i$ and $b_i$ are randomly drawn from uniform distribution on $[0,1]$. 
The training is performed by calculating optimal weight $\bm{w}^{(l)}$ for each $l$ by Eq.~\eqref{eq:optimal-weight}.
We define the output $\hat{l}$ from the trained model by 
\begin{align}
    \hat{l} = \mathrm{argmax}_l \{{\bm{w}^{(l)}}^{\mathrm{T}}\bm{\phi}(\bm{x})\}.
\end{align}

\subsection{Baseline}
We first obtain the baseline to compare with the mod-FKV algorithm by performing exact singular value decomposition of the matrix $\bm{X}$.
We implement this by \texttt{lstsq} function available within NumPy package \cite{harris2020array}.

\begin{figure}[t]
    \centering{
        \includegraphics[width=\linewidth]{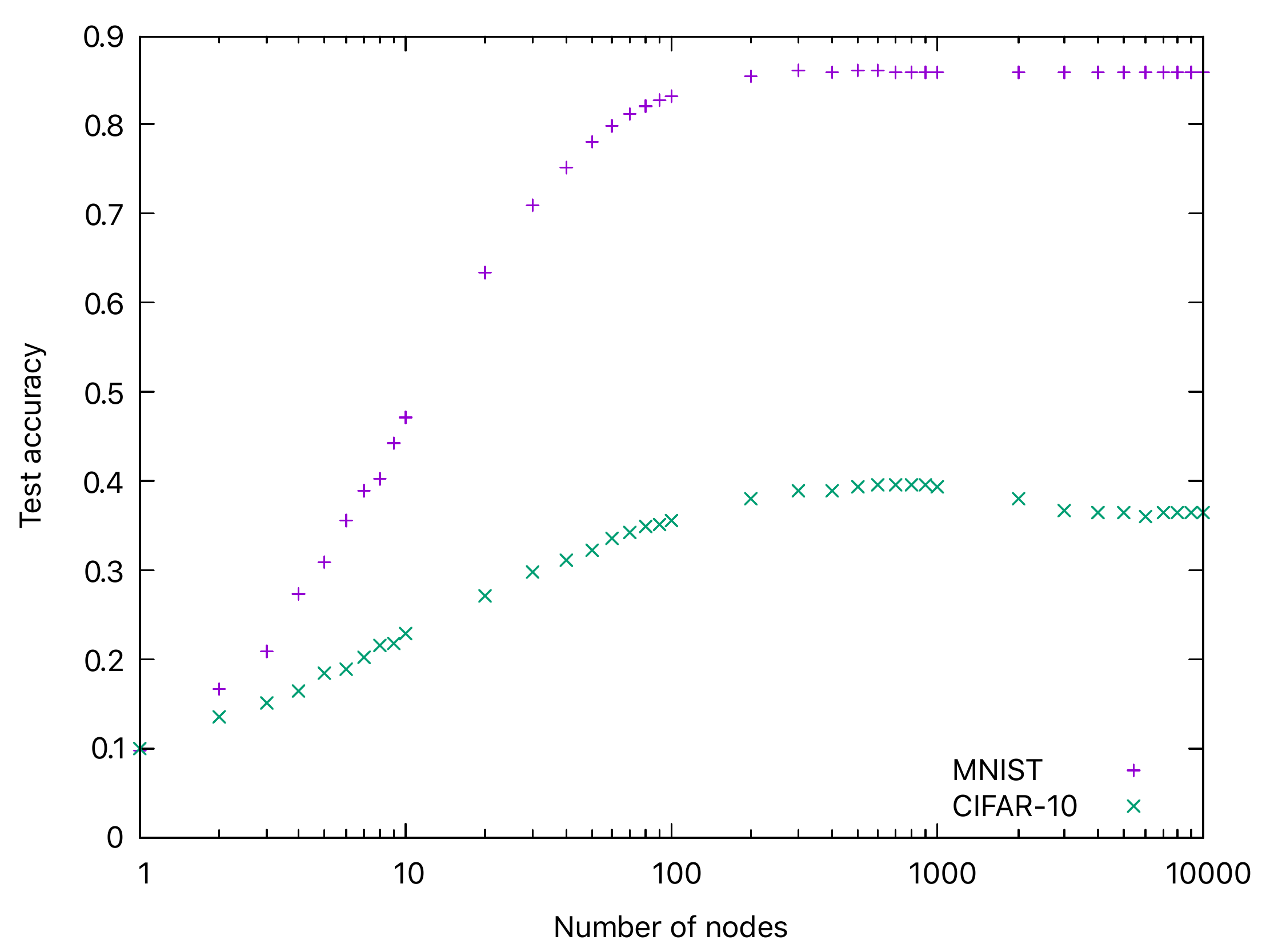}
        \caption{Change of test accuracy with varying number of nodes using the exact signular value decomposition with \texttt{lstsq}. \label{fig:nodes_accuracy}}
    }
\end{figure}

First, we observe how the test accuracy varies with respect to the number of nodes $M$.
Figure~\ref{fig:nodes_accuracy} shows the numerical result.
For the MNIST dataset, we see that the test accuracy increases as we enlarge $M$ until it saturates around 85\% at around $M=200$.
For the CIFAR-10, the best performances are obtained when $M$ is around 1000, and it degrades for larger $M$.
Although the performances of the extreme learning for both of the datasets are far from the state-of-the-art, note that we are not interested in achieving it.
The advantage of extreme learning is that its training is extremely simpler and faster compared to neural networks with complicated structures.
The result presented in Fig.~\ref{fig:nodes_accuracy} constitutes the baseline for all of the following numerical experiments; it is the best possible performance that one can hope for using the extreme learning approach.

\begin{figure}[t]
    \centering{
        \includegraphics[width=\linewidth]{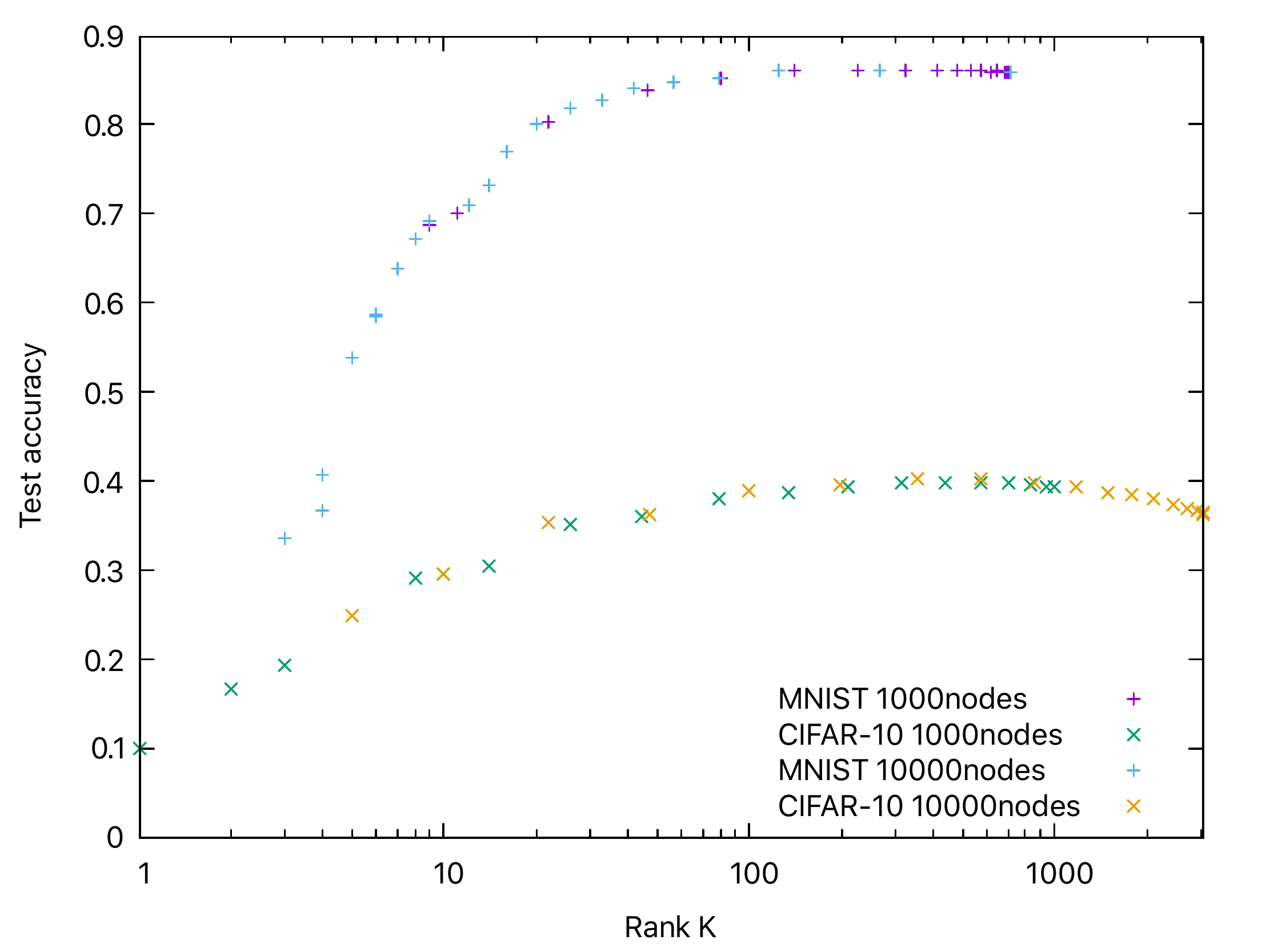}
        \caption{Effect of low-rank approximation to the test accuracy using \texttt{lstsq}. \label{fig:mnist_ranks}}
    }
\end{figure}

Next, we observe how the test accuracy behaves when we use $\bm{X}_K^{+}$ instead of $\bm{X}^+$ to compute $\bm{w}$.
Choosing $M=10^3$ and $10^4$, we vary $K$ to see changes in the test accuracy.
The result is shown in Fig.~\ref{fig:mnist_ranks}.
From the figure, we observe that the same value of $K$ results in the same level of accuracy even for different $M$. 
In particular, a low-rank approximation can recover the performance degradation of the $M=10^4$ case; it gave us lower test accuracy when we used the exact pseudo-inverse $\bm{X}^+$ compared to $M=10^3$ (see Fig.~\ref{fig:nodes_accuracy}).
Also, for any integer $n$, the test accuracy of $(M,K)=(10^4,n)$ case is roughly larger than that of $M=n$ with exact pseudo-inverse case (Fig.~\ref{fig:nodes_accuracy}).
For example, test accuracy of $(M, K)=(10^4,10)$ is about 0.7 while that of $M=10$ with exact pseudo inverse is about 0.4.

These results motivate us to use the following strategy for constructing an extreme learning model: use as large $M$ as possible to generate random features and then truncate the singular value at a certain level.
This strategy is particularly suited for the mod-FKV algorithm since the larger $M$ implies the larger difference in the computational cost between conventional algorithms and the quantum-inspired one.

\subsection{Quantum-inspired algorithm}
We implement the quantum-inspired algorithm in Sec.~\ref{sec:quantum-inspired} on Python and compare its performance with the conventional approach that uses \texttt{lstsq}.
More concretely, we compare the computational time required for the mod-FKV($K, P$) algorithm to achieve the comparable test accuracy with respect to the $K$-rank approximation performed by \texttt{lstsq}.
In the following numerical experiments, we fix the rank to $K=10$ and the number of nodes to $M=10^3$ or $10^4$ which achieve about 70\% and 30\% test accuracy respectively for MNIST and CIFAR-10 datasets with \texttt{lstsq}.

We also investigate a possible shortcoming of this application that, because the connections $\bm{a}_i$ of the neural network are taken randomly, the elements of the matrix $\bm{X}$ might have a rather uniform distribution.
Therefore, one might concern that the Frobenius norm sampling according to Eqs. \eqref{eq:fi} and \eqref{eq:gj} can be replaced by uniform sampling without degrading the performance.
To investigate this concern, we conduct the following experiment.
We perform the classification of two datasets by extreme learning, where pseudo-inverses of the matrix $\bm{X}$ are obtained with mod-FKV($10, P$) as presented in Sec. \ref{sec:quantum-inspired} and a modified version of mod-FKV($10,P$) that uses uniform sampling instead of Eqs. \eqref{eq:fi} and \eqref{eq:gj}.

\begin{figure}[t]
    \centering{
        \includegraphics[width=\linewidth]{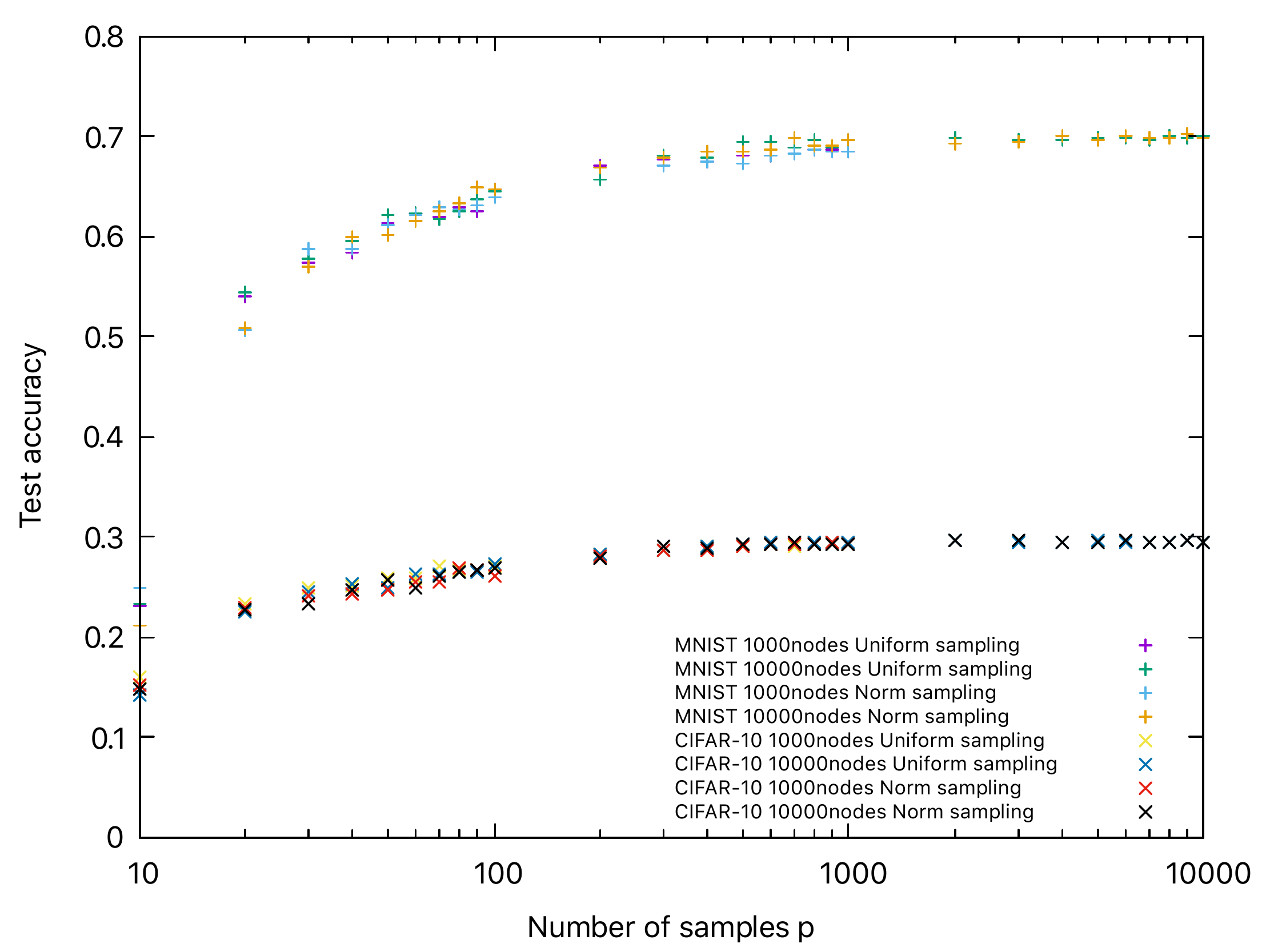}
        \caption{Change of test accuracy with respect to the number of samples $P$ used in the mod-FKV algorithm. \label{fig:samples-accuracy}}
    }
\end{figure}

First, we observe how many samples are needed to obtain the comparable test accuracy to \texttt{lstsq} by varying the number of samples, $P$.
Figure \ref{fig:samples-accuracy} shows how the test accuracy of the mod-FKV-based extreme learning improves with increasing $P$.
From the figure, we see that taking $P=10^2$ achieves an accuracy of 65\% and 25\% respectively for MNIST and CIFAR-10 datasets, which is slightly worse than 70\% and 30\% achieved by \texttt{lstsq} but comparable.
We, therefore, compare the computational time of mod-FKV-based extreme learning with that of the \texttt{lstsq}-based one afterward with $P=10^2$.

Table~\ref{table:time_ac_norm_uniform_sampling} shows a comparison of the test accuracy and the time required for training (time to obtain $\bm{w}$) with each methods at a fixed number of samples $P=10^2$.
We achieve the shortest computational time by using uniform sampling approach for both of the $M=10^3$ and $M=10^4$ cases.
For the Frobenius-norm sampling approach, the computational time is reduced from \texttt{lstsq} only for the $M=10^4$ case.
This is because of the overhead required for constructing the segment-tree data structure, which becomes less significant when $M$ is large.
The test accuracy of both sampling strategies are about 64\% to 65\% and at the same level.
Therefore, our concern that the Frobenius-norm sampling can be replaced by uniform sampling seems to be correct, that is, the Frobenius norm sampling utilized in the mod-FKV algorithm is not effective in this setting and the uniform sampling gives us an equivalent test accuracy.
This means the quantum-inspired algorithm is not effective in general for naive extreme learning.

\begin{table*}
    \centering{
    \caption{Comparison of test accuracy and computational time of the extreme learning with \texttt{lstsq}, mod-FKV with Frobenius norm and uniform sampling applied for MNIST. For the Frobenius sampling, we show the time required for constructing the segment-tree data structure separately. \label{table:time_ac_norm_uniform_sampling}}
    \begin{ruledtabular}
        \begin{tabular}{llll}
            $M$ & Method & Test accuracy & Training time [s] \\
            \(10^3\) & \texttt{lstsq} (Rank-10 approximation) & \(0.687\pm0.00626\) & 1.00 \\
            \(10^3\) & mod-FKV($10, 10^2$) (Norm sampling) & \(0.640\pm0.0158\) & $1.96+0.09$ \\
            \(10^3\) & mod-FKV($10, 10^2$) (Uniform sampling) & \(0.646\pm0.0241\) & 0.0555 \\
            \(10^4\) & \texttt{lstsq} (Rank-10 approximation) & \(0.693\pm0.00362\) & 105 \\
            \(10^4\) & mod-FKV($10, 10^2$) (Norm sampling) & \(0.648\pm0.0143\) & $9.53+0.67$ \\
            \(10^4\) & mod-FKV($10, 10^2$) (Uniform sampling) & \(0.644\pm0.0213\) & 0.535
        \end{tabular}
    \end{ruledtabular}
    }
\end{table*}

Note that, even though the uniform sampling approach is significantly faster than other methods, the Frobenius norm sampling is also order-of-magnitude faster than \texttt{lstsq} when $M$ is large (see the $M=10^4$ case in Table \ref{table:time_ac_norm_uniform_sampling}).
To see the advantage of the quantum-inspired algorithm, the above observation implies that we need to apply the mod-FKV to a more ``concentrated'' matrix.
We, therefore, train the weights $\bm{a}_i$ after the first optimization of the output weight vector $\bm{w}$.
More concretely, we use the following algorithm:
\begin{enumerate}
    \item Set random $\bm{a}_i$ and $b_i$ and calculate Eq. \eqref{eq:approx-pseudo-inverse} to obtain an optimal weight $\bm{w}_1^{(l)}$.
    \item Train $\bm{a}_i$ and $b_i$ while fixing $\bm{w}$ to the one obtained in the previous step. Training is performed to minimize the squared loss defined by 
    \begin{align}
        \sum_{i=1}^D \sum_{l=1}^{10} (y_i^{(l)}-{\bm{w}^{(l)}_1}^{\mathrm{T}}\bm{\phi}(\bm{x}_i))^2. 
    \end{align}
    Let the trained values $\bm{a}_i^*$ and $b_i^*$.
    \item Using $\bm{a}_i^*$ and $b_i^*$, calculate Eq. \eqref{eq:approx-pseudo-inverse} again to obtain an optimal weight $\bm{w}_2^{(l)}$.
    \item Use $\bm{a}_i^*$, $b_i^*$ and $\bm{w}_2^{(l)}$ to evaluate the test accuracy.
\end{enumerate}
We expect this procedure to make the matrix $\bm{X} = (\bm{\phi}(\bm{x}_1)~\bm{\phi}(\bm{x}_2)~\cdots~\bm{\phi}(\bm{x}_D))$ somewhat non-uniform.

\begin{figure}[t]
    \centering{
        \includegraphics[width=\linewidth]{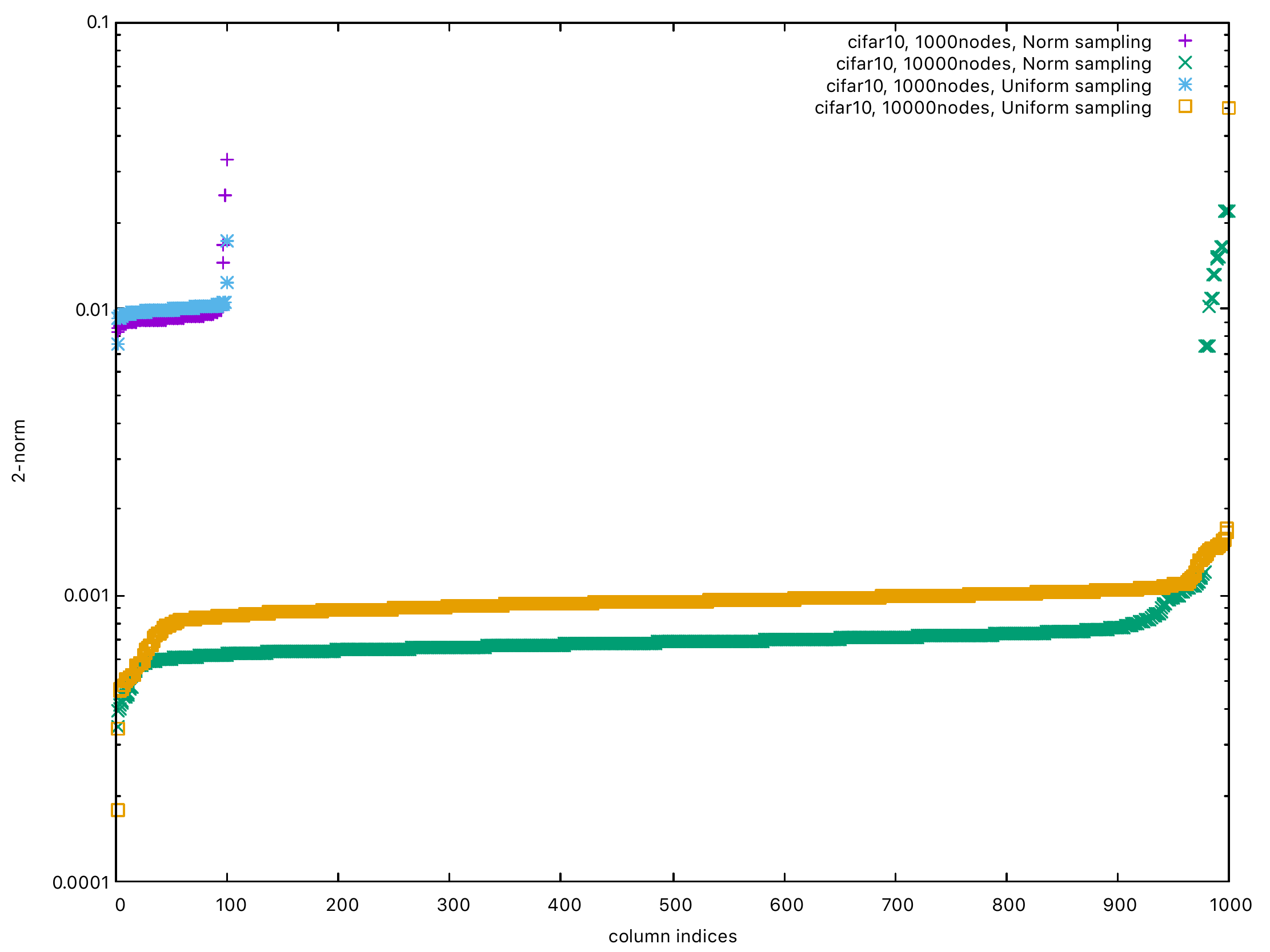}
        \caption{Norms of columns sampled by different strategies after the optimization of $\bm{a}_i$ and $b$. \label{fig:cifar10_selected_norm}}
    }
\end{figure}

Now, we show the effect of this optimization as Table \ref{table:optimized_weights_test_acc_MNIST} and \ref{table:optimized_weights_test_acc_CIFAR}.
We find that, in the $M=10^4$ case, this treatment makes the Frobenius norm sampling effective as expected.
This is because, through the optimization of $\bm{a}_i$ and $b_i$, the matrix $\bm{X}$ have become non-uniform.
On the other hand, we find it to be ineffective for the $M=10^3$ case.
To see why this happens, we further analyze the norm of the columns sampled in each sampling strategy, which is shown in Fig. \ref{fig:cifar10_selected_norm}.
From the figure, we see that the norms of sampled columns does not vary very much for $M=10^3$ with respect to different strategies, but they do for $M=10^4$.
This is because the optimized $\bm{a}_i^*$ in $M=10^4$ case gives more non-uniform column norms. 
This result indicates that we should apply the quantum-inspired algorithm when the target matrix is rather non-uniform because, otherwise, the use of uniform sampling is sufficient to obtain comparable results.

\begin{table}
    \centering{
    \caption{Comparison of test accuracy after the optimization of parameters $\bm{a}_i$ and $b_i$ with Frobenius norm and uniform sampling applied for MNIST. \label{table:optimized_weights_test_acc_MNIST}}
    \begin{ruledtabular}
        \begin{tabular}{lll}
            $M$ & Method & Test accuracy \\ 
            \(10^3\) & mod-FKV($10, 10^2$) (Norm sampling) & \(0.704\pm0.0707\) \\
            \(10^3\) & mod-FKV($10, 10^2$) (Uniform sampling) & \(0.584\pm0.0771\) \\
            \(10^4\) & mod-FKV($10, 10^3$) (Norm sampling) & \(0.844\pm0.0136\) \\
            \(10^4\) & mod-FKV($10, 10^3$) (Uniform sampling) & \(0.745\pm0.0406\)
        \end{tabular}
    \end{ruledtabular}
    }
\end{table}

\begin{table}
    \centering{
    \caption{Comparison of test accuracy after the optimization of parameters $\bm{a}_i$ and $b_i$ with Frobenius norm and uniform sampling applied for CIFAR-10. \label{table:optimized_weights_test_acc_CIFAR}}
    \begin{ruledtabular}
        \begin{tabular}{lll}
            $M$ & Method & Test accuracy \\ 
            \(10^3\) & mod-FKV($10, 10^2$) (Norm sampling) & \(0.224\pm0.0175\) \\
            \(10^3\) & mod-FKV($10, 10^2$) (Uniform sampling) & \(0.236\pm0.0196\) \\
            \(10^4\) & mod-FKV($10, 10^3$) (Norm sampling) & \(0.309\pm0.00957\) \\
            \(10^4\) & mod-FKV($10, 10^3$) (Uniform sampling) & \(0.233\pm0.0177\)
        \end{tabular}
    \end{ruledtabular}
    }
\end{table}

\section{Conclusion}
We apply the quantum-inspired algorithm to a machine learning framework called extreme learning.
We find that mod-FKV algorithm is effective in reducing the time required for training.
Even though the implementation of the mod-FKV algorithm in this work is not quite optimized, it achieves considerable speedup compared to the efficient NumPy implementation of the exact singular value decomposition.
However, an important observation is that the Frobenius norm sampling, which is the core of the quantum-inspired singular value decomposition, is not always required.
Our experiments indicate that, under certain circumstances when the elements of target matrix is uniform, it is better to use naive uniform sampling.
On the other hand, when the matrix is non-uniform, we find the Frobenius norm sampling is effective to quickly compute a low-rank approximation.

A few possible future directions are in order.
First, our implementation of mod-FKV algorithm is not optimized for speed.
For example, rewriting in C should improve the runtime by a constant factor.
Second, it is not clear if we can benefit from the Frobenius sampling in concrete examples of other tasks such as recommendation systems, which the original quantum-inspired algorithm is designed for.
We should be aware of distribution of matrix elements.
Finally, it would be interesting to look for other application of the mod-FKV algorithm, given that it can reduce the computational time to certain extent for the application explored in this work.
For example, we are looking into a possibility of its application to the reservoir computing approach \cite{jaeger2004harnessing}, which is a method to learn temporal datasets.

\begin{acknowledgements}
KM is supported by JST PRESTO Grant No. JPMJPR2019.
This work is supported by MEXT Quantum Leap Flagship Program (MEXTQLEAP) Grant No. JPMXS0118067394 and JPMXS0120319794. We also acknowledge support from JST COI-NEXT program Grant No. JPMJPF2014.
\end{acknowledgements}

\bibliography{90}

\end{document}